\documentclass[lettersize,journal]{IEEEtran}
\usepackage{amsmath,amsfonts}
\usepackage{algorithmic}
\usepackage{algorithm}
\usepackage{array}
\usepackage[caption=false,font=normalsize,labelfont=sf,textfont=sf]{subfig}
\usepackage{textcomp}
\usepackage{stfloats}
\usepackage{url}
\usepackage{verbatim}
\usepackage{graphicx}
\usepackage{cite}
\usepackage{booktabs}
\usepackage{balance}
\usepackage{multirow}
\graphicspath{{figures/}}
\begin{document}

\title{Execution-Feasible Dispatch of LLM Pre-training Loads for Renewable-Dominant AI Data Centers}

\author{Xi Tong, Ershun Du*,~\IEEEmembership{Member,~IEEE}, Haiyang Jiang,~\IEEEmembership{Member,~IEEE}, Long Chen,~\IEEEmembership{Member,~IEEE}, Qin Wang*,~\IEEEmembership{Senior Member,~IEEE}%

\thanks{Corresponding authors: Ershun Du and Qin Wang.}

\thanks{X. Tong and Q. Wang are with the Department of Electrical and Electronic
Engineering, The Hong Kong Polytechnic University, Hong Kong SAR, China.(e-mail:polyu-eee-xi.tong@connect.polyu.hk, qin-ee.wang@polyu.edu.hk.)}
\thanks{E. Du is with the Lab of Low Carbon Energy, Tsinghua University, Beijing, China.(e-mail:duershun@tsinghua.edu.cn)}
\thanks{H. Jiang is with the Department of Electrical Engineering, Tsinghua University, Beijing, China.(e-mail:jiang\_hy@tsinghua.edu.cn)} 
\thanks{L. Chen is with Tsinghua University and the Department of Computing, The Hong Kong Polytechnic University, China.(e-mail:longchen.cs@ieee.org)}}

\markboth{IEEE Transactions on Smart Grid,~Vol.~XX, No.~X, XXX~2026}%
{Tong: Execution-Feasible Dispatch of LLM Pre-training Loads for AIDCs}

\maketitle

\begin{abstract}
AI data centers (AIDCs) are emerging as large power-system loads, making their operational flexibility increasingly valuable. Existing studies often treat AIDC demand as continuously adjustable or generically delay-tolerant, which overstating the flexibility of large language model (LLM) pre-training. This paper proposes an execution-feasible dispatch framework for LLM pre-training in renewable-dominant AIDCs.  LLM descriptors are first converted into checkpoint-domain daily workloads, then mapped to feasible 3D-parallel GPU modes with explicit throughput, GPU-capacity, memory, and IT-power limits. The resulting checkpoint- and mode-constrained load states are embedded into a day-ahead power dispatch model. Case studies on a 2048 GPUs AIDC with two LLaMA pre-training tasks show that checkpoint-aware 3D dispatch reduces operating cost by 20.02\% and renewable curtailment from 1.077 to 0.014~MWh relative to fixed continuous training. Related results also show that checkpoint records alone do not create load flexibility,practical grid-interactive flexibility arises from checkpoint waiting combined with execution-feasible 3D-mode compensation.
\end{abstract}


\section{Introduction}

\subsection{Background and Motivation}
\IEEEPARstart{A}{rtificial} intelligence (AI) data centers are becoming new class of large electric loads in modern power systems. The rapid deployment of large language models (LLMs), foundation models, and GPU-accelerated services is increasing both the scale and the temporal concentration of the demand for data-center electricity \cite{Masanet2020Science,Koomey2021Joule,deVries2023Joule,Patterson2022Computer}. Meanwhile, high renewable penetration requires more flexibility on the demand-side to mitigate renewable uncertainty, curtailment, and peak-stress conditions. Recent studies indicate that AI data centers (AIDCs) can potentially act as grid-interactive assets through software-defined workload orchestration, demand response, and power scheduling \cite{Colangelo2025NatureEnergy,Lin2024Eenergy}. Therefore, AIDCs should not be modeled only as passive high-power consumers, their computing-side flexibility should be converted into power system readable load states.

However, this conversion is nontrivial for LLM pre-training. Most existing data center load models describe computing demand as aggregate power representations or workload-shifting abstractions \cite{Qureshi2009SIGCOMM,Liu2013PE,Ghamkhari2013TSG,Ma2024IoTJ}. Recent studies have further refined data center load flexibility through optimal energy consumption characteristics, service level agreements (SLAs), or geographically dispersed operation \cite{Zhou2025AE,Basmadjian2017TSG,Kiani2018TSG,Pahlevan2021TSC}. These abstractions are useful for conventional Internet services and cloud batch jobs, but they may overestimate the flexibility of LLM pre-training. LLM pre-training is a sequential optimization process composed of a large number of training iterations. Its temporal shift is constrained by checkpoint states, while its resumption power level is determined by discrete feasible GPU parallel configurations rather than by a continuous power command \cite{Shoeybi2019Megatron,Narayanan2021Megatron}. In addition, memory feasibility, GPU capacity, parallel efficiency, and deadline-constrained training progress jointly determine whether delayed workload can be recovered. Thus, the key challenge is not simply to shift AIDC demand, but to derive an execution-feasible AIDC load model that can participate in power dispatch.

\subsection{Literature Review}
The energy footprint and grid impact of data centers have been widely studied. Global data center energy analyzes show that efficiency improvements have historically mitigated a part of demand growth, while the rapid expansion of cloud services and AI accelerators still makes data centers important electricity consumers \cite{Masanet2020Science,Koomey2021Joule,deVries2023Joule}. For AI workloads, the energy and carbon implications of training large neural models and foundation models have been reported in recent studies \cite{Strubell2019ACL,Patterson2022Computer,Luccioni2022BLOOM}. These works motivate low-carbon and grid-aware AIDC operation, but they mainly quantify energy or carbon impacts and do not provide execution-feasible load models for power system dispatch.

Data center demand response and green operation have also been investigated from the perspectives of workload shifting, local generation, microgrid operation, and renewable-powered operation \cite{Liu2013PE,Liu2015TON,LiQi2018TSG,Goiri2013ASPLOS}. Other studies consider electricity cost optimization in green data centers \cite{Yang2018AE,Basmadjian2017TSG,Kiani2018TSG,Pahlevan2021TSC}. Integrated energy-system studies further couple data centers with smart grids, storage, and on-site renewable generation \cite{Guo2021AE,Long2023TSTE,Guo2014TPDS,Cao2018TON}. More recent studies refine data center flexibility by considering carbon-aware computing, QoS-assured demand response, and dispatchable capacity evaluation \cite{Radovanovic2023TPS,Zhang2022TSC,Vafamehr2019TSG,Han2026TSG}. Multi-data-center coordination has also been studied for tie-line power smoothing and cross-site operation \cite{Yang2024TCC}. These studies establish data centers as useful demand-side resources for cost reduction, peak management, and renewable absorption. Nevertheless, most formulations still rely on aggregate flexible-demand envelopes, divisible workload models, or utilization-based power models. Such representations cannot explicitly determine when an LLM training load can be safely paused, what discrete power state it will take after resumption, and whether the delayed progress can be completed before the deadline.

The computer systems literature provides the missing execution mechanisms of LLM training. Large-scale Transformer training is enabled by tensor parallelism, pipeline parallelism and data parallelism\cite{Shoeybi2019Megatron,Narayanan2021Megatron,GPipe2019,PipeDream2019}. Memory pressure is addressed through optimizer-state partitioning, sublinear-memory checkpointing, and tensor re-materialization \cite{ZeRO2020,Chen2016Checkpoint,Jain2020Checkmate}. Activation recomputation is further refined for large Transformer training \cite{Korthikanti2023Reducing}. Recent carbon-aware machine learning (ML) cluster schedulers also show that job shifting and resource scaling can reduce carbon footprint or peak power at the cluster level \cite{Xu2025GREEN}. However, these techniques are mainly designed for throughput, memory efficiency, or cluster workload orchestration. Current research on the interaction between LLM pre-training and power system is still scarce.

\subsection{Contributions and Paper Organization}
Although current literature offers robust strategies for workload shifting and multi-site coordination, these general approaches do not directly translate into an execution-feasible load model for LLM pre-training.
In particular, these approaches often overlook three parts: the safe temporal boundaries for load interruption, the discrete power profiles upon task resumption, and the feasibility of deadline-constrained progress recovery.
To fill this gap, this paper develops an execution-feasible AIDC dispatch framework that converts checkpoint states and feasible three-dimensional (3D) parallel GPU modes into the day-ahead power dispatch model. 
The main contributions are summarized as follows.
\begin{itemize}
\item An execution-feasible LLM pre-training load model is developed by integrating checkpoint progress and feasible 3D-parallel GPU modes.
\item A checkpoint-state graph based dispatch method is proposed to preserve GPU, memory, throughput, and deadline feasibility during load shifting.
\item The system value of execution-feasible LLM load flexibility is quantified in a renewable-dominant AIDC, revealing its cost-saving, curtailment-reduction, and BESS-complementary effects.
\end{itemize}

The remainder of this paper is organized as follows. Section II introduces the AIDC system framework and the flexibility mechanism of LLM pre-training loads. Section III presents the execution-feasible load modeling method. Section IV formulates the day-ahead power dispatch model. Section V reports the numerical case study, and Section VI concludes the paper.

\section{System Framework and Problem Description}

\subsection{Framework of Green-Power Direct Supply AIDC System}
\begin{figure}[!t]
\centering
\includegraphics[width=\columnwidth]{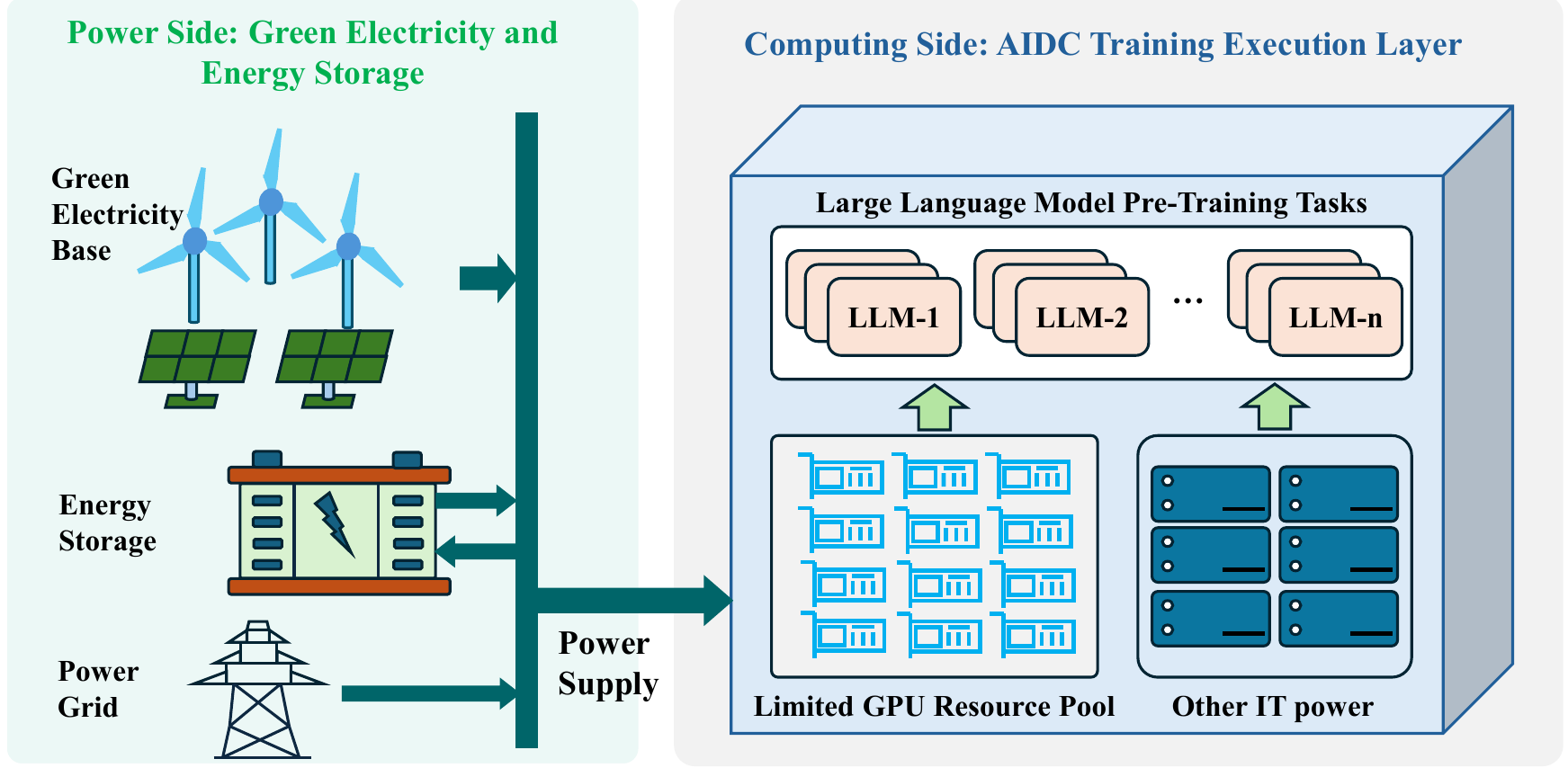}
\caption{Architecture of the green-power direct-supply AIDC system.}
\label{fig:aidc_architecture}
\end{figure}

Fig.~\ref{fig:aidc_architecture} shows the considered green-power direct-supply AIDC. Local wind and photovoltaic generation are preferentially delivered to the AIDC through a dedicated or equivalent green-power interface, while the upstream grid provides reliability support and time-of-use price signals. Unlike islanded operation, renewable generation, grid purchase, curtailment, and BESS operation are jointly coordinated to satisfy AIDC demand \cite{LiQi2018TSG,Goiri2013ASPLOS}.

On the computing side, the AIDC electric load consists of non-schedulable base IT power and schedulable LLM pre-training power. The external electric load is obtained from the internal IT power through the power usage effectiveness (PUE) \cite{Barroso2019Datacenter}:
\begin{equation}
P_t^{\mathrm{AIDC}}
=
\mathrm{PUE}
\left(
P_t^{\mathrm{base,IT}}
+
P_t^{\mathrm{train,IT}}
\right),
\quad \forall t \in \mathcal{T},
\label{eq:aidc_load}
\end{equation}
where $P_t^{\mathrm{base,IT}}$ is the basic IT power, and $P_t^{\mathrm{train,IT}}$ is the power of active LLM pre-training tasks.

The green-power bus balance is
\begin{equation}
P_t^{\mathrm{ren,av}}
-
P_t^{\mathrm{cur}}
+
P_t^{\mathrm{grid}}
+
P_t^{\mathrm{dis}}
=
P_t^{\mathrm{AIDC}}
+
P_t^{\mathrm{ch}},
\quad \forall t \in \mathcal{T},
\label{eq:power_balance_interface}
\end{equation}
where $P_t^{\mathrm{ren,av}}$, $P_t^{\mathrm{cur}}$, $P_t^{\mathrm{grid}}$, $P_t^{\mathrm{ch}}$, and $P_t^{\mathrm{dis}}$ are available renewable power, renewable curtailment, grid purchase, BESS charging, and BESS discharging, respectively. 

\subsection{Flexibility Description of LLM Pre-Training Loads}

\begin{figure}[!t]
\centering
\includegraphics[width=\columnwidth]{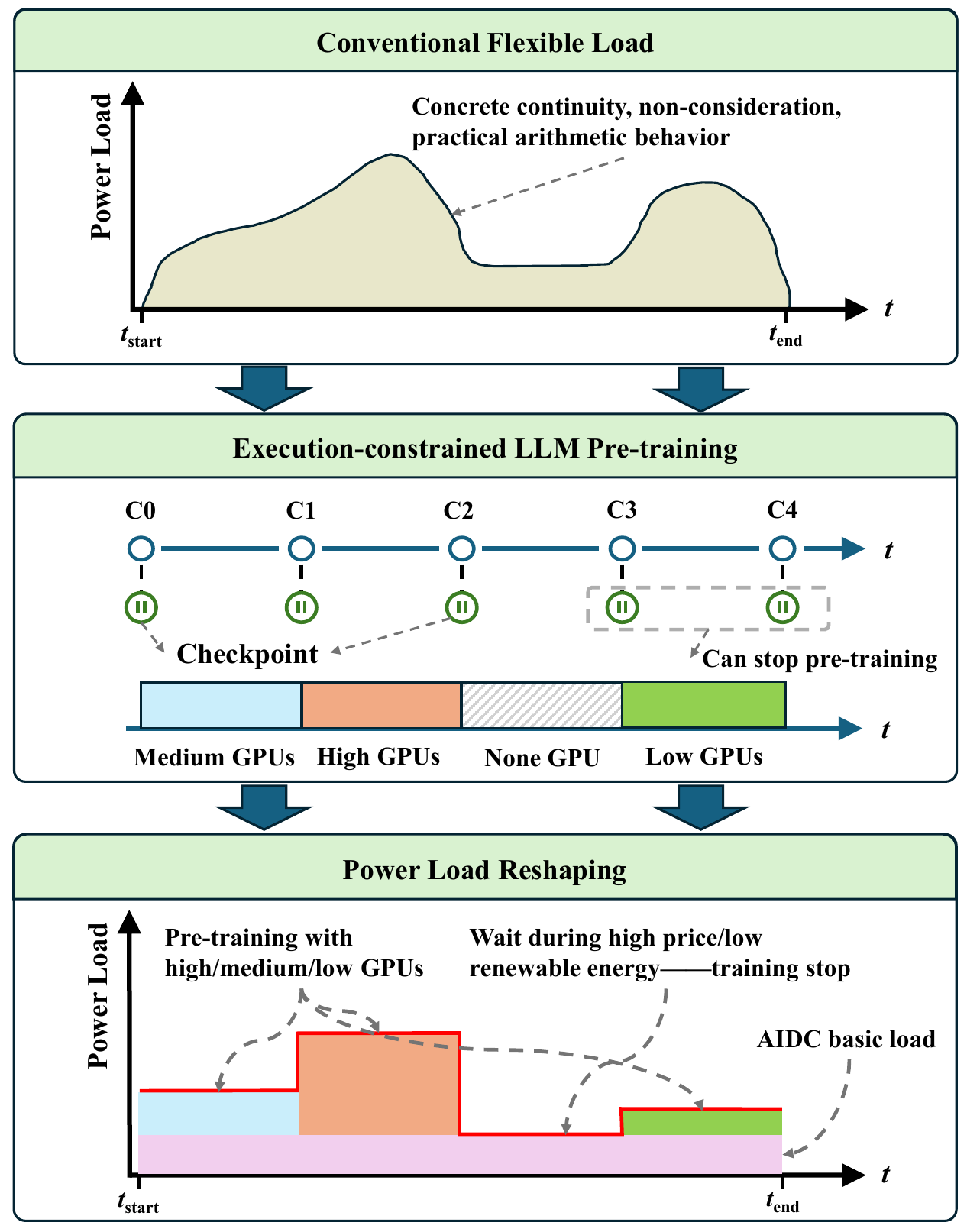}
\caption{Conceptual flexibility of LLM pre-training loads for power dispatch.}
\label{fig:flexibility_concept}
\end{figure}

Fig.~\ref{fig:flexibility_concept} contrasts conventional flexible loads with LLM pre-training loads. Conventional data-center models often use continuously adjustable power, delay-tolerant jobs, or utilization-based envelopes \cite{Ghamkhari2013TSG,Ma2024IoTJ}, while LLM pre-training can safely wait or resume only at checkpoint states. Waiting reduces instantaneous training power but does not reduce remaining computation. After resumption, the power level is discrete and determined by feasible GPU execution modes involving tensor, pipeline, and data parallelism, memory feasibility, throughput, and GPU capacity \cite{Shoeybi2019Megatron,Narayanan2021Megatron}. Thus, LLM pre-training flexibility is a set of checkpoint-constrained and mode-discrete execution states, where checkpoints provide temporal boundaries and feasible 3D modes provide recovery capability \cite{Chen2016Checkpoint,Korthikanti2023Reducing}. The task--computing--resource mapping chain is
\begin{equation}
\Omega_j^{\mathrm{task}}
\longrightarrow
{C_j,\mathcal{S}_j^{\mathrm{ckpt}}}
\longrightarrow
\mathcal{A}_j^{\mathrm{eff}}
\longrightarrow
P_t^{\mathrm{train,IT}}
\longrightarrow
P_t^{\mathrm{AIDC}},
\label{eq:tcr_chain}
\end{equation}
where $\Omega_j^{\mathrm{task}}$ is the LLM task descriptor, $C_j$ is the daily training workload, $\mathcal{S}_j^{\mathrm{ckpt}}$ is the checkpoint-domain progress set, and $\mathcal{A}_j^{\mathrm{eff}}$ is the pruned feasible 3D-parallel mode set.

\subsection{Coordination Framework and Dispatch Problem}

\begin{figure*}[!t]
\centering
\includegraphics[width=\textwidth]{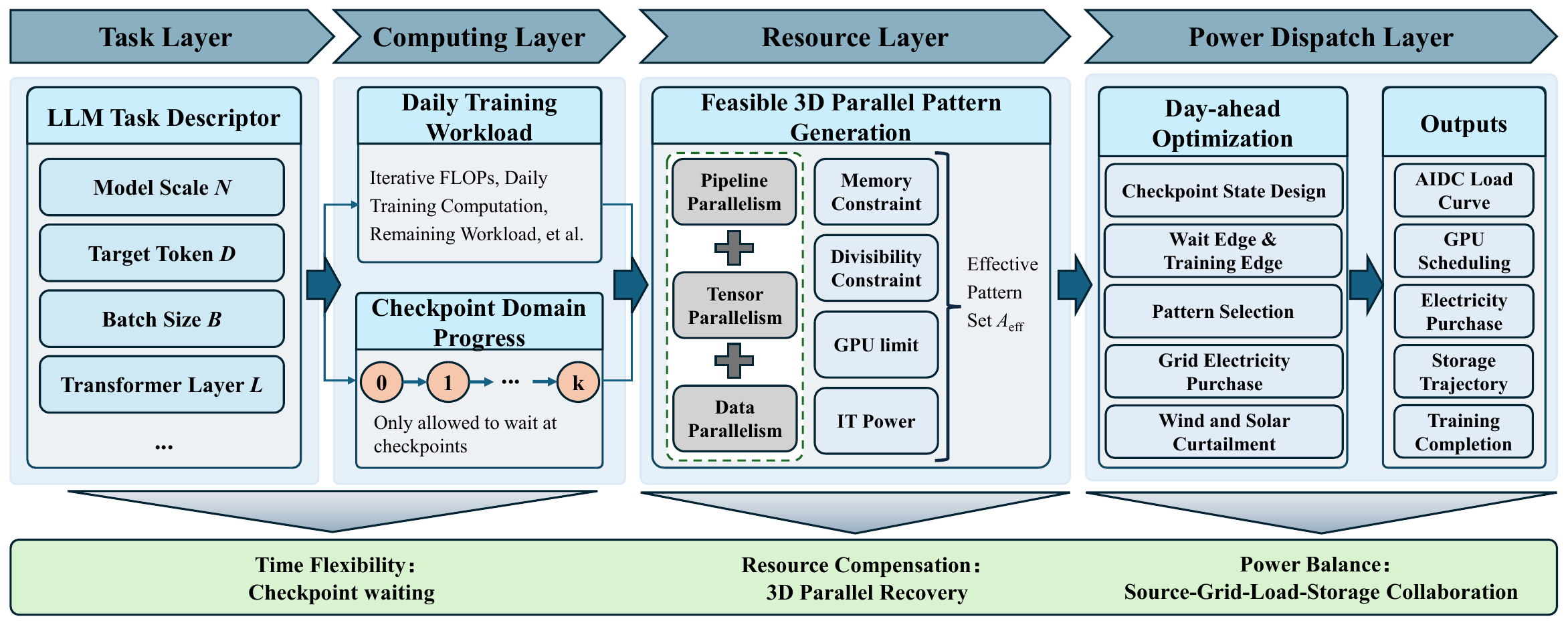}
\caption{Task--computing--resource coordination framework.}
\label{fig:coordination_framework}
\end{figure*}

Based on this mechanism, Fig.~\ref{fig:coordination_framework} gives the proposed coordination framework. The task layer describes model size, token target, batch size, sequence length, and deadline. The computing layer converts these descriptors into daily workload and checkpoint progress, and the resource layer maps them into feasible 3D GPU modes with different throughput, GPU occupation, memory usage, and IT power. The dispatch layer then coordinates these execution-feasible load states with renewable generation, grid purchase, curtailment, and BESS operation under checkpoint-state continuity, completion, GPU-capacity, AIDC-capacity, BESS, and power-balance constraints.

\section{Modeling of LLM Pre-training Loads}
\label{sec:tcr_modeling}

\subsection{LLM Training Workflow and Task Descriptor}
\label{subsec:workflow_descriptor}

\begin{figure}[!t]
\centering
\includegraphics[width=\columnwidth]{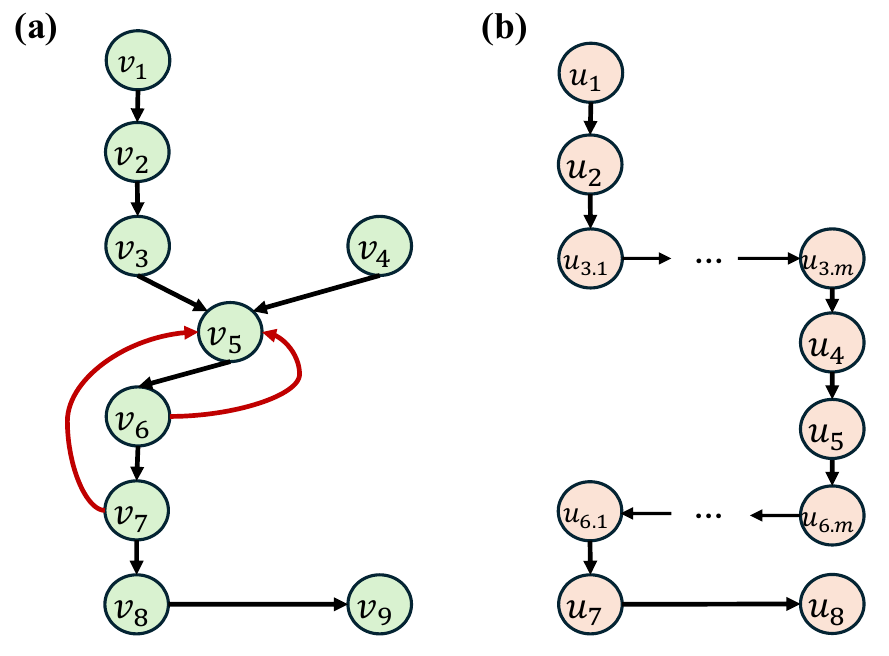}
\caption{Workflow of (a) LLM training and (b) Iterative training.}
\label{fig:llm_workflow}
\end{figure}

Fig.~\ref{fig:llm_workflow} shows the macro-level LLM workflow and the iteration-level distributed training loop. The proposed model focuses on iterative pre-training and periodic checkpointing; data preparation, validation, model export, and post-training stages are not scheduled. Forward/backward computation and distributed communication determine workload and feasible execution modes, while checkpoints provide safe pause/resume states for power dispatch \cite{Korthikanti2023Reducing}.

For each LLM pre-training job $j\in\mathcal{J}$, the task descriptor is defined as
\begin{equation}
\Omega_j^{\mathrm{task}}
=
\left\{
N_j,D_j,B_j,s_j,l_j,h_j,h_{f,j},n_{h,j},V_j,r_j,T_j^{\mathrm{dead}}
\right\},
\label{eq:task_descriptor}
\end{equation}
where $N_j$, $D_j$, $B_j$, and $s_j$ are model size, full token target, global batch size, and sequence length; $l_j$, $h_j$, $h_{f,j}$, $n_{h,j}$, and $V_j$ are Transformer architecture parameters; $r_j$ is the daily tranche ratio; and $T_j^{\mathrm{dead}}$ is the deadline.

\subsection{Task-to-Computing Mapping and Checkpoint-Domain Progress}
\label{subsec:task_to_computing}

For a dense causal-decoder Transformer, the per-iteration training FLOPs can be approximated according to commonly used Transformer FLOPs accounting in LLM training \cite{Narayanan2021Megatron}. The per-iteration workload is written as
\begin{equation}
F_{\mathrm{iter},j}
=
\chi_j^{\mathrm{F}}
B_j s_j l_j h_j^2
\left(
1+\frac{s_j}{6h_j}+\frac{V_j}{16l_jh_j}
\right),
\label{eq:fiter}
\end{equation}
where $\chi_j^{\mathrm{F}}$ is the training FLOPs coefficient; the three terms in parentheses capture dense matrix operations, attention-related computation, and vocabulary projection. The coefficient can absorb FLOPs-counting conventions and implementation overheads.

The number of full-task iterations and the number of daily-tranche iterations are
\begin{subequations}
\label{eq:iteration_numbers}
\begin{align}
I_j &= \left\lceil \frac{D_j}{B_j s_j}\right\rceil,
\label{eq:full_iter}\\
I_j^{\mathrm{day}} &= \left\lceil r_j I_j\right\rceil.
\label{eq:day_iter}
\end{align}
\end{subequations}
Accordingly, the daily-tranche computing workload is
\begin{equation}
C_j=I_j^{\mathrm{day}}F_{\mathrm{iter},j}.
\label{eq:daily_workload}
\end{equation}
Let $S_{j,t}\in[0,1]$ be the normalized progress of the daily tranche at time interval $t$. The remaining computing workload is
\begin{equation}
C_{j,t}^{\mathrm{rem}}=(1-S_{j,t})C_j.
\label{eq:remaining_workload}
\end{equation}

To represent safe waiting states, the daily tranche is divided into $K_j^{\mathrm{seg}}$ aggregated checkpoint-to-checkpoint segments. The checkpoint-domain index set and progress set are
\begin{equation}
\mathcal{K}_j=\left\{0,1,\ldots,K_j^{\mathrm{seg}}\right\},\quad
\mathcal{S}_j^{\mathrm{ckpt}}
=
\left\{
\frac{k}{K_j^{\mathrm{seg}}}\;\middle|\; k\in\mathcal{K}_j
\right\}.
\label{eq:ckpt_set}
\end{equation}
The workload of one aggregated checkpoint segment is
\begin{equation}
C_j^{\mathrm{seg}}=\frac{C_j}{K_j^{\mathrm{seg}}}.
\label{eq:segment_workload}
\end{equation}
Checkpoint waiting produces no training progress, but the required computation is conserved:
\begin{equation}
\sum_{t\in\mathcal{T}}X_{j,t}\Delta t=C_j,
\label{eq:workload_conservation}
\end{equation}
where $X_{j,t}$ is the realized training throughput in FLOPs/s. Checkpointing does not reduce total computation; it only enables temporal reshaping if delayed workload can be recovered by a faster feasible mode.

\subsection{Computing-to-Resource Mapping and 3D-Parallel Execution Mode}
\label{subsec:computing_to_resource}

\begin{figure}[!t]
\centering
\includegraphics[width=\columnwidth]{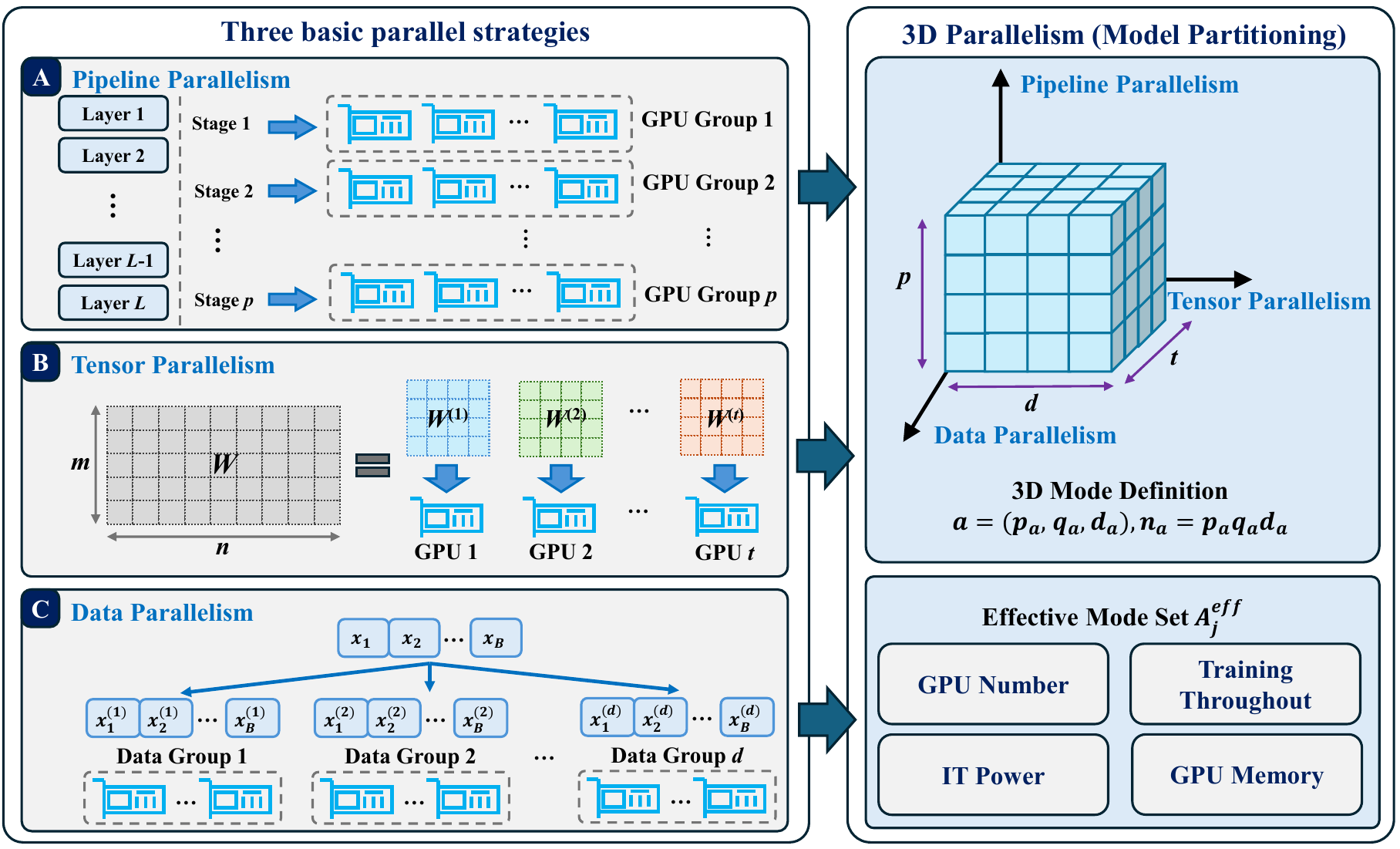}
\caption{Basic parallel strategies, 3D-parallel mode definition, and effective resource attributes for LLM training.}
\label{fig:tcr_mapping}
\end{figure}

Fig.~\ref{fig:tcr_mapping} illustrates the computing-to-resource mapping. Pipeline, tensor, and data parallelism jointly form a 3D execution mode, which determines GPU occupation, throughput, and training IT power.

A candidate 3D-parallel configuration is defined as
\begin{subequations}
\label{eq:3d_config}
\begin{align}
a&=(p_a,q_a,d_a),
\label{eq:3d_tuple}\\
n_a&=p_a q_a d_a,
\label{eq:gpu_number}
\end{align}
\end{subequations}
where $p_a$, $q_a$, and $d_a$ are the pipeline-, tensor-, and data-parallel sizes; $n_a$ is the required number of GPUs. The 3D mode links model partitioning with GPU occupation and training power in the AIDC dispatch model.

Given the per-GPU peak FLOP rate $R_g$ and the end-to-end efficiency $\eta_{j,a}$, the effective training throughput and the compute amount within one scheduling interval are
\begin{equation}
X_{j,a}=n_aR_g\eta_{j,a},\quad
\bar{X}_{j,a}=X_{j,a}\Delta t.
\label{eq:throughput}
\end{equation}
where $X_{j,a}$ is in FLOPs/s; $\bar{X}_{j,a}$ is in FLOPs per interval; and $\eta_{j,a}$ is the efficiency factor, which can be obtained from cluster profiling records or calibrated performance models; in the numerical study, it is used as a representative mode attribute to capture pipeline bubbles, communication, synchronization, and implementation losses.

Memory feasibility is screened by
\begin{subequations}
\label{eq:memory_model}
\begin{align}
M_{j,a}^{\mathrm{state}}
&=
\frac{\omega_s N_j}{p_aq_a\zeta_a^{\mathrm{state}}},
\label{eq:state_mem}\\
M_{j,a}^{\mathrm{act}}
&=
\rho_j^{\mathrm{rec}}\frac{l_j}{p_a}
\Bigg(
\lambda_h\frac{B_j}{d_a}s_j\frac{h_j}{q_a}
+\lambda_f\frac{B_j}{d_a}s_j\frac{h_{f,j}}{q_a}
\notag\\
&\qquad\qquad
+\lambda_a\frac{B_j}{d_a}s_j^2\frac{n_{h,j}}{q_a}
\Bigg)
\notag\\
&\quad
+\lambda_o\frac{B_j}{d_a}s_j\frac{h_j}{q_a}
+\lambda_v\frac{B_j}{d_a}s_j\frac{V_j}{q_a},
\label{eq:act_mem}\\
M_{j,a}^{\mathrm{GPU}}
&=
M_{j,a}^{\mathrm{state}}
+M_{j,a}^{\mathrm{act}}
+M_0
+\delta^{\mathrm{other}}
\left(M_{j,a}^{\mathrm{state}}+M_{j,a}^{\mathrm{act}}\right).
\label{eq:gpu_mem}
\end{align}
\end{subequations}
where $\omega_s$ is the mixed-precision model-state memory coefficient; $\zeta_a^{\mathrm{state}}$ is the state-partitioning factor, with $\zeta_a^{\mathrm{state}}=1$ if no state partitioning is used \cite{ZeRO2020}; $\rho_j^{\mathrm{rec}}$ is the activation-recomputation coefficient; $\lambda_h$, $\lambda_f$, $\lambda_a$, $\lambda_o$, and $\lambda_v$ are calibrated activation coefficients; and $M_0$ and $\delta^{\mathrm{other}}$ are fixed and proportional runtime memory overheads.

A mode is memory-feasible only if
\begin{equation}
M_{j,a}^{\mathrm{GPU}}\le M_g^{\mathrm{cap}}.
\label{eq:memory_feasible}
\end{equation}
In addition to memory feasibility, 3D-parallel execution must satisfy engineering divisibility conditions:
\begin{subequations}
\label{eq:divisibility}
\begin{align}
\operatorname{mod}(l_j,p_a)&=0,
\label{eq:layer_div}\\
\operatorname{mod}(h_j,q_a)&=0,\quad
\operatorname{mod}(h_{f,j},q_a)=0,
\label{eq:hidden_div}\\
\operatorname{mod}(n_{h,j},q_a)&=0,\quad
\operatorname{mod}(B_j,d_a)=0.
\label{eq:head_batch_div}
\end{align}
\end{subequations}
These constraints ensure valid layer assignment, tensor/head partitioning, and batch splitting.

The IT power of mode $a$ is modeled as
\begin{equation}
P_{j,a}^{\mathrm{IT}}
=
n_aP_g^{\mathrm{train}}+P_{j,a}^{\mathrm{comm}},
\label{eq:mode_power}
\end{equation}
where $P_g^{\mathrm{train}}$ is the equivalent active training power per GPU; $P_{j,a}^{\mathrm{comm}}$ is the communication and mode-dependent overhead. In the case study, $P_{j,a}^{\mathrm{comm}}$ is absorbed into the equivalent mode power coefficient.

\subsection{Feasible Mode Set and Dominance Pruning}
\label{subsec:mode_pruning}

For each job $j$, the feasible 3D-parallel mode set is defined as
\begin{equation}
\begin{aligned}
\mathcal{A}_j=\bigl\{a\ \bigm|\ &
M_{j,a}^{\mathrm{GPU}}\le M_g^{\mathrm{cap}},\quad
n_a\le N_{\mathrm{GPU}}^{\max},\\
&\eqref{eq:divisibility}\ \text{is satisfied}
\bigr\}.
\end{aligned}
\label{eq:feasible_mode_set}
\end{equation}
Each feasible mode is defined by
\begin{equation}
\phi_{j,a}
=
\left\{
n_a,X_{j,a},\bar{X}_{j,a},P_{j,a}^{\mathrm{IT}},M_{j,a}^{\mathrm{GPU}}
\right\}.
\label{eq:mode_tuple}
\end{equation}
The tuple $\phi_{j,a}$ gives the GPU occupation, throughput, interval workload, IT power, and memory requirement of one executable mode.

To reduce the number of online dispatch choices, a dominance pruning rule is applied. Mode $a$ is dominated by mode $b$ if
\begin{equation}
n_b\le n_a,\quad
P_{j,b}^{\mathrm{IT}}\le P_{j,a}^{\mathrm{IT}},\quad
X_{j,b}\ge X_{j,a},
\label{eq:dominance}
\end{equation}
with at least one strict inequality. A dominated mode consumes no less GPU resource and IT power while providing no higher throughput, and can be removed. The remaining effective mode set is $\mathcal{A}_j^{\mathrm{eff}}$. These modes are execution-feasible candidates under the adopted hardware and model assumptions, rather than proprietary production configurations of a specific LLM training system. Algorithm~\ref{alg:mode_pruning} summarizes their generation and pruning.

\begin{algorithm}[!t]
\footnotesize
\caption{Feasible 3D-Parallel Configuration Generation and Pruning}
\label{alg:mode_pruning}
\begin{algorithmic}[1]
\REQUIRE Task descriptor $\Omega_j^{\mathrm{task}}$; GPU parameters $R_g$ and $M_g^{\mathrm{cap}}$; candidate sets $\mathcal{P}$, $\mathcal{Q}$, and $\mathcal{D}$; GPU limit $N_{\mathrm{GPU}}^{\max}$.
\ENSURE Effective mode set $\mathcal{A}_j^{\mathrm{eff}}$.
\STATE Initialize $\mathcal{A}_j\leftarrow\emptyset$.
\FORALL{$p\in\mathcal{P}$, $q\in\mathcal{Q}$, and $d\in\mathcal{D}$}
    \STATE Set $a=(p,q,d)$ and $n_a=pqd$.
    \IF{the divisibility conditions in \eqref{eq:divisibility} are violated}
        \STATE Continue to the next candidate.
    \ENDIF
    \STATE Compute $X_{j,a}$ and $\bar{X}_{j,a}$ according to \eqref{eq:throughput}.
    \STATE Compute $M_{j,a}^{\mathrm{GPU}}$ according to \eqref{eq:memory_model}.
    \STATE Compute $P_{j,a}^{\mathrm{IT}}$ according to \eqref{eq:mode_power}.
    \IF{$M_{j,a}^{\mathrm{GPU}}\le M_g^{\mathrm{cap}}$ and $n_a\le N_{\mathrm{GPU}}^{\max}$}
        \STATE Add $a$ and $\phi_{j,a}$ to $\mathcal{A}_j$.
    \ENDIF
\ENDFOR
\STATE Remove dominated modes from $\mathcal{A}_j$ according to \eqref{eq:dominance}.
\STATE Return the remaining modes as $\mathcal{A}_j^{\mathrm{eff}}$.
\end{algorithmic}
\end{algorithm}

\section{Checkpoint-Aware Day-Ahead Green-Power Dispatching Model}
\label{sec:dispatch_model}

\subsection{Mode-Pruned Checkpoint-State Graph}

\begin{figure}[!t]
\centering
\includegraphics[width=\columnwidth]{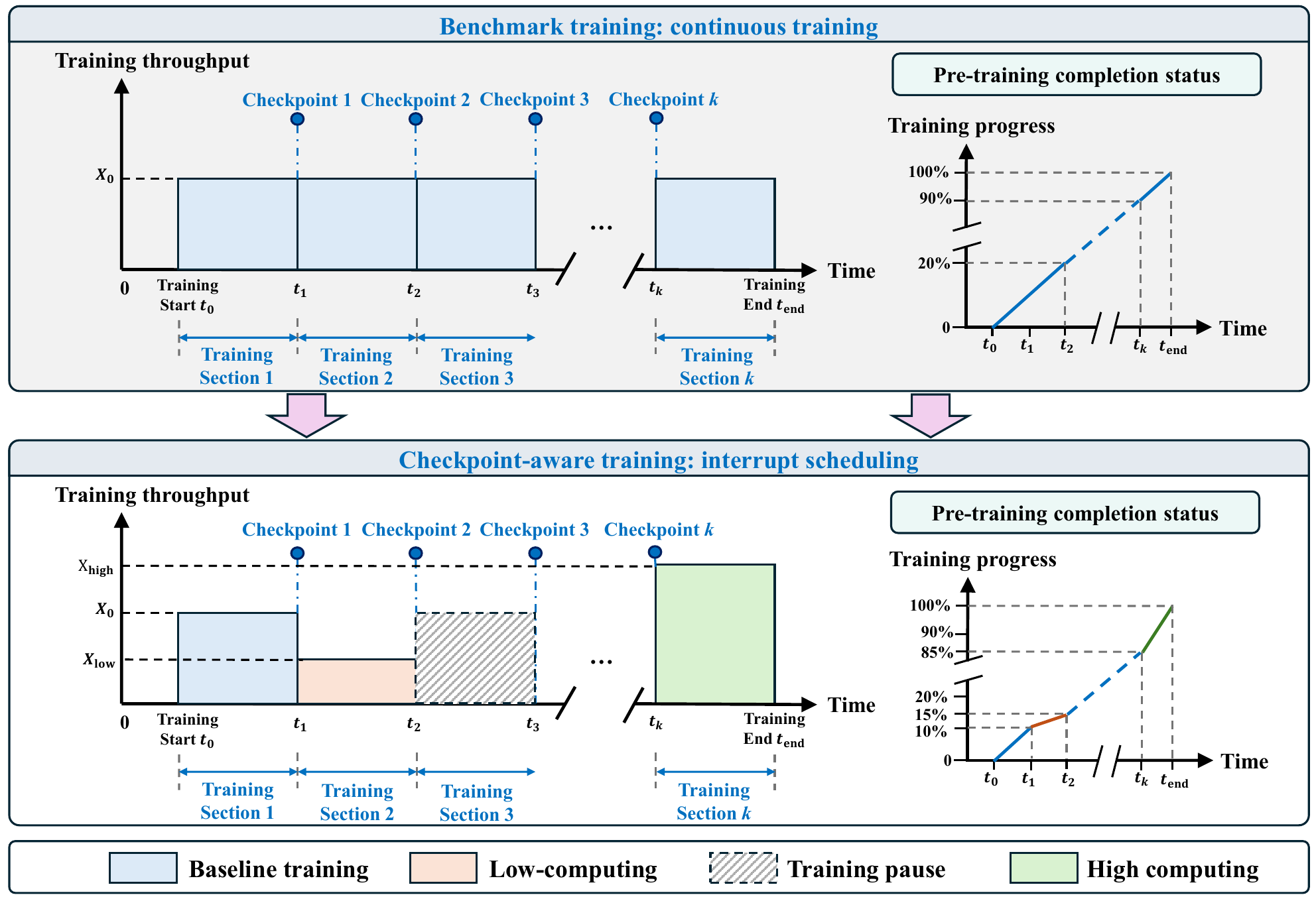}
\caption{Checkpoint-aware interruptible dispatch mechanism compared with benchmark continuous training.}
\label{fig:ckpt_3d_mechanism}
\end{figure}
Fig.~\ref{fig:ckpt_3d_mechanism} compares fixed continuous training with checkpoint-aware dispatch. In the benchmark case, the job is executed with a fixed throughput $X_0$. In the proposed case, state changes occur only at checkpoint boundaries; low/high-computing segments correspond to feasible 3D modes in $\mathcal{A}_j^{\mathrm{eff}}$, and pauses correspond to checkpoint waiting. Incompatible mode transitions can be excluded during edge generation.

Let $\mathcal{T}=\{1,2,\ldots,T\}$ be the set of dispatch intervals and $\mathcal{H}=\{0,1,\ldots,T\}$ be the set of time-state nodes. For each training job $j$, a checkpoint-state graph $\mathcal{G}_j=(\mathcal{V}_j,\mathcal{E}_j)$ is constructed as
\begin{equation}
\mathcal{V}_j=
\left\{(\tau,k)\mid \tau\in\mathcal{H},\ k\in\mathcal{K}_j\right\}
\cup \{g_j\},
\label{eq:graph_nodes}
\end{equation}
where $(\tau,k)$ is the state in which job $j$ has reached checkpoint $k$ at the beginning of time state $\tau$, and $g_j$ is an artificial sink node. The source node is $s_j=(0,0)$.

Two types of edges are introduced. The first type is the waiting edge:
\begin{equation}
e:(\tau,k)\rightarrow(\tau+1,k),
\quad \tau=0,\ldots,T-1,
\label{eq:waiting_edge}
\end{equation}
which means that job $j$ waits for one dispatch interval at checkpoint $k$.

The second type is the training edge:
\begin{equation}
e:(\tau,k)\rightarrow
(\tau+\ell_{j,a,\Delta k},k+\Delta k),
\label{eq:training_edge}
\end{equation}
where $a\in\mathcal{A}_j^{\mathrm{eff}}$ is the selected 3D-parallel mode, $\Delta k$ is the number of checkpoint segments continuously executed by this edge, $1\le \Delta k\le K^{\mathrm{jump}}$, and $k+\Delta k\le K_j^{\mathrm{seg}}$. The parameter $K^{\mathrm{jump}}$ is the maximum number of checkpoint segments represented by one training edge.

For a training edge $e$, the executed workload and integer duration are
\begin{subequations}
\label{eq:edge_duration}
\begin{align}
C_{j,e}^{\mathrm{edge}}
&=
\frac{\Delta k}{K_j^{\mathrm{seg}}}C_j,
\label{eq:edge_workload}\\
\ell_{j,a,\Delta k}
&=
\left\lceil
\frac{C_{j,e}^{\mathrm{edge}}}{\bar{X}_{j,a}}
\right\rceil .
\label{eq:edge_length}
\end{align}
\end{subequations}
The active slot set of edge $e$ is
\begin{equation}
\Gamma_e=
\{\tau+1,\tau+2,\ldots,\tau+\ell_{j,a,\Delta k}\}.
\label{eq:active_slot_set}
\end{equation}
A training edge is generated only if
$\tau+\ell_{j,a,\Delta k}\le T_j^{\mathrm{dead}}$.
Finally, zero-power completion edges connect every terminal checkpoint node $(\tau,K_j^{\mathrm{seg}})$ with $\tau\le T_j^{\mathrm{dead}}$ to the sink node $g_j$.

Let $x_{j,e}\in\{0,1\}$ indicate whether edge $e$ is selected. The selected edges must form one feasible path from $s_j$ to $g_j$. The path constraints are written in a unified net-flow form:
\begin{subequations}
\label{eq:path_constraints}
\begin{align}
\sum_{e\in\delta_j^+(s_j)}x_{j,e}
-
\sum_{e\in\delta_j^-(s_j)}x_{j,e}
&=1,
\label{eq:source_flow}\\
\sum_{e\in\delta_j^+(v)}x_{j,e}
-
\sum_{e\in\delta_j^-(v)}x_{j,e}
&=0,
\quad \forall v\in\mathcal{V}_j\setminus\{s_j,g_j\},
\label{eq:middle_flow}\\
\sum_{e\in\delta_j^-(g_j)}x_{j,e}
-
\sum_{e\in\delta_j^+(g_j)}x_{j,e}
&=1.
\label{eq:sink_flow}
\end{align}
\end{subequations}
where $\delta_j^+(v)$ and $\delta_j^-(v)$ are the outgoing and incoming edge sets of node $v$, respectively. Constraints~\eqref{eq:path_constraints} enforce checkpoint-state continuity and daily-tranche completion. Since waiting edges only start from checkpoint nodes, non-checkpoint interruption is excluded by graph construction.

The ceiling operation in \eqref{eq:edge_length} conservatively counts the last partially utilized slot as occupied, avoiding underestimation of the time and power required to complete a checkpoint segment.

\subsection{Edge-to-Power Decoding and Resource Constraints}

For each selected training edge, let $a(e)$ be the associated 3D-parallel mode and let $\Delta k(e)$ be the number of checkpoint segments completed by the edge. The mode-use indicator is linearly decoded as
\begin{equation}
y_{j,a,t}
=
\sum_{\substack{
e\in\mathcal{E}_j^{\mathrm{tr}}:\\
a(e)=a,\ t\in\Gamma_e}}
x_{j,e},
\quad \forall j,a,t,
\label{eq:mode_use}
\end{equation}
where $\mathcal{E}_j^{\mathrm{tr}}$ is the set of training edges. Because the selected path is time-monotonic, at most one training edge can be active for a given job in a given time interval.

The aggregate training IT power and GPU usage are
\begin{subequations}
\label{eq:training_power_gpu}
\begin{align}
P_t^{\mathrm{train,IT}}
&=
\sum_{j\in\mathcal{J}}
\sum_{a\in\mathcal{A}_j^{\mathrm{eff}}}
P_{j,a}^{\mathrm{IT}}y_{j,a,t},
\label{eq:train_power}\\
G_t^{\mathrm{train}}
&=
\sum_{j\in\mathcal{J}}
\sum_{a\in\mathcal{A}_j^{\mathrm{eff}}}
n_a y_{j,a,t}.
\label{eq:gpu_usage}
\end{align}
\end{subequations}
Eq.~\eqref{eq:train_power} gives the schedulable training IT power formed by active GPU modes, while \eqref{eq:gpu_usage} gives the corresponding GPU occupation. Together with the non-schedulable base IT power, the total AIDC electric load is
\begin{equation}
P_t^{\mathrm{AIDC}}
=
\mathrm{PUE}
\left(
P_t^{\mathrm{base,IT}}+P_t^{\mathrm{train,IT}}
\right),
\quad \forall t\in\mathcal{T}.
\label{eq:aidc_power}
\end{equation}
Thus, selected edges determine active 3D modes, active modes determine training IT power, and PUE converts IT power into total AIDC load.

The GPU-capacity and AIDC power-capacity constraints are
\begin{subequations}
\label{eq:resource_constraints}
\begin{align}
G_t^{\mathrm{train}}
&\le N_{\mathrm{GPU}}^{\max},
\quad \forall t\in\mathcal{T},
\label{eq:gpu_capacity}\\
P_t^{\mathrm{AIDC}}
&\le P^{\mathrm{AIDC,max}},
\quad \forall t\in\mathcal{T}.
\label{eq:power_capacity}
\end{align}
\end{subequations}
For reporting purposes, the normalized progress of job $j$ by the end of interval $t$ can be decoded from the completed training edges:
\begin{equation}
S_{j,t}
=
\frac{1}{K_j^{\mathrm{seg}}}
\sum_{\substack{
e\in\mathcal{E}_j^{\mathrm{tr}}:\\
\bar{\tau}(e)\le t}}
\Delta k(e)x_{j,e},
\quad \forall t\in\mathcal{T},
\label{eq:progress_decode}
\end{equation}
where $\bar{\tau}(e)$ is the ending time state of edge $e$. This decoded progress variable is not required for feasibility because the path constraints already ensure that each job reaches the terminal checkpoint before its deadline.

\subsection{Green-Power Bus and BESS Operation Constraints}

The power-balance interface introduced in Section~II is imposed as an operational constraint at the green-power bus:
\begin{equation}
P_t^{\mathrm{ren,av}}
-
P_t^{\mathrm{cur}}
+
P_t^{\mathrm{grid}}
+
P_t^{\mathrm{dis}}
=
P_t^{\mathrm{AIDC}}
+
P_t^{\mathrm{ch}},
\quad \forall t\in\mathcal{T}.
\label{eq:power_balance}
\end{equation}
Renewable curtailment and grid purchase are bounded by
\begin{subequations}
\label{eq:grid_ren_limits}
\begin{align}
0\le P_t^{\mathrm{cur}}
&\le P_t^{\mathrm{ren,av}},
\quad \forall t\in\mathcal{T},
\label{eq:curtail_limit}\\
0\le P_t^{\mathrm{grid}}
&\le P^{\mathrm{grid,max}},
\quad \forall t\in\mathcal{T}.
\label{eq:grid_limit}
\end{align}
\end{subequations}

The BESS state is indexed on $\mathcal{H}$, while the charging and discharging powers are indexed on $\mathcal{T}$. The BESS dynamics, power limits, SOC limits, and charge/discharge exclusiveness are
\begin{subequations}
\label{eq:bess_constraints}
\begin{align}
E_t^{\mathrm{bat}}
&=
E_{t-1}^{\mathrm{bat}}
+
\eta^{\mathrm{ch}}P_t^{\mathrm{ch}}\Delta t
-
\frac{P_t^{\mathrm{dis}}}{\eta^{\mathrm{dis}}}\Delta t,
\quad \forall t\in\mathcal{T},
\label{eq:bess_soc}\\
0\le P_t^{\mathrm{ch}}
&\le u_t^{\mathrm{ch}}P^{\mathrm{ch,max}},
\quad \forall t\in\mathcal{T},
\label{eq:bess_ch_power}\\
0\le P_t^{\mathrm{dis}}
&\le u_t^{\mathrm{dis}}P^{\mathrm{dis,max}},
\quad \forall t\in\mathcal{T},
\label{eq:bess_dis_power}\\
u_t^{\mathrm{ch}}+u_t^{\mathrm{dis}}
&\le 1,
\quad u_t^{\mathrm{ch}},u_t^{\mathrm{dis}}\in\{0,1\},
\quad \forall t\in\mathcal{T},
\label{eq:bess_binary}\\
E^{\min}
\le E_t^{\mathrm{bat}}
&\le E^{\max},
\quad \forall t\in\mathcal{H},
\label{eq:bess_energy_limits}\\
E_T^{\mathrm{bat}}
&=
E_0^{\mathrm{bat}}.
\label{eq:bess_cyclic}
\end{align}
\end{subequations}
For no-BESS scenarios, $P_t^{\mathrm{ch}}=P_t^{\mathrm{dis}}=0$, and the BESS state equations are removed.

\subsection{Objective Function and Solution Procedure}

The day-ahead dispatch objective is
\begin{equation}
\begin{aligned}
\min J
=&
\sum_{t\in\mathcal{T}}
\pi_t^{\mathrm{grid}}P_t^{\mathrm{grid}}\Delta t
+
\sum_{t\in\mathcal{T}}
c^{\mathrm{cur}}P_t^{\mathrm{cur}}\Delta t  \\
&+
\sum_{t\in\mathcal{T}}
c^{\mathrm{bat}}
\left(P_t^{\mathrm{ch}}+P_t^{\mathrm{dis}}\right)\Delta t .
\end{aligned}
\label{eq:objective}
\end{equation}
The three terms are upstream-grid purchase cost, renewable-curtailment opportunity cost, and BESS throughput cost. The GPU pool is assumed to be a dedicated pre-training pool, so the opportunity cost of reallocating GPUs to external services is not included in the base objective.

The complete optimization model is
\begin{equation}
\begin{aligned}
\min\quad
& J \\
\mathrm{s.t.}\quad
& \eqref{eq:path_constraints},\\
& \eqref{eq:mode_use}-\eqref{eq:resource_constraints},\\
& \eqref{eq:power_balance}-\eqref{eq:bess_constraints},\\
& x_{j,e}\in\{0,1\},\quad
u_t^{\mathrm{ch}},u_t^{\mathrm{dis}}\in\{0,1\},\\
& P_t^{\mathrm{grid}},P_t^{\mathrm{cur}},
P_t^{\mathrm{ch}},P_t^{\mathrm{dis}},E_t^{\mathrm{bat}}\ge 0.
\end{aligned}
\label{eq:complete_milp}
\end{equation}
The model is a MILP. Memory, divisibility, and 3D-mode engineering feasibility are embedded in $\mathcal{A}_j^{\mathrm{eff}}$ by Algorithm~\ref{alg:mode_pruning}, while $y_{j,a,t}$, $P_t^{\mathrm{train,IT}}$, $G_t^{\mathrm{train}}$, and $P_t^{\mathrm{AIDC}}$ are linearly decoded from selected checkpoint-state paths. Algorithm~\ref{alg:dayahead_tcr} summarizes the procedure.

\begin{algorithm}[!t]
\footnotesize
\caption{Checkpoint-Aware Training Graph Construction and Dispatch}
\label{alg:dayahead_tcr}
\begin{algorithmic}[1]
\REQUIRE $\{\Omega_j^{\mathrm{task}}\}$, $\{\mathcal{A}_j^{\mathrm{eff}}\}$, $\{K_j^{\mathrm{seg}}\}$, renewable and price forecasts, BESS and AIDC parameters.
\ENSURE Training paths, 3D-mode schedules, AIDC load, grid purchase, curtailment, and BESS operation.
\FORALL{training job $j\in\mathcal{J}$}
    \STATE Initialize the checkpoint-state graph $\mathcal{G}_j=(\mathcal{V}_j,\mathcal{E}_j)$ by \eqref{eq:graph_nodes}.
    \STATE Add checkpoint waiting edges by \eqref{eq:waiting_edge}.
    \STATE Generate feasible training edges using $\mathcal{A}_j^{\mathrm{eff}}$, $\Delta k\le K^{\mathrm{jump}}$, and \eqref{eq:training_edge}--\eqref{eq:active_slot_set}.
    \STATE Attach edge attributes $\{a(e),\Delta k(e),\Gamma_e,n_{a(e)},P_{j,a(e)}^{\mathrm{IT}}\}$ and remove infeasible edges violating deadlines.
    \STATE Add zero-power completion edges to the sink node $g_j$.
\ENDFOR
\STATE Construct the path-selection MILP with \eqref{eq:path_constraints}.
\STATE Decode selected edges into training power, GPU usage, and AIDC load by \eqref{eq:mode_use}--\eqref{eq:aidc_power}.
\STATE Add resource, green-power bus, and BESS constraints \eqref{eq:resource_constraints}--\eqref{eq:bess_constraints}.
\STATE Solve \eqref{eq:complete_milp} and recover training paths, progress, load, grid purchase, curtailment, and BESS trajectories.
\end{algorithmic}
\end{algorithm}

\section{Numerical Case Study}

\subsection{Basic Data and Case Settings}

\begin{figure}[!t]
\centering
\includegraphics[width=\columnwidth]{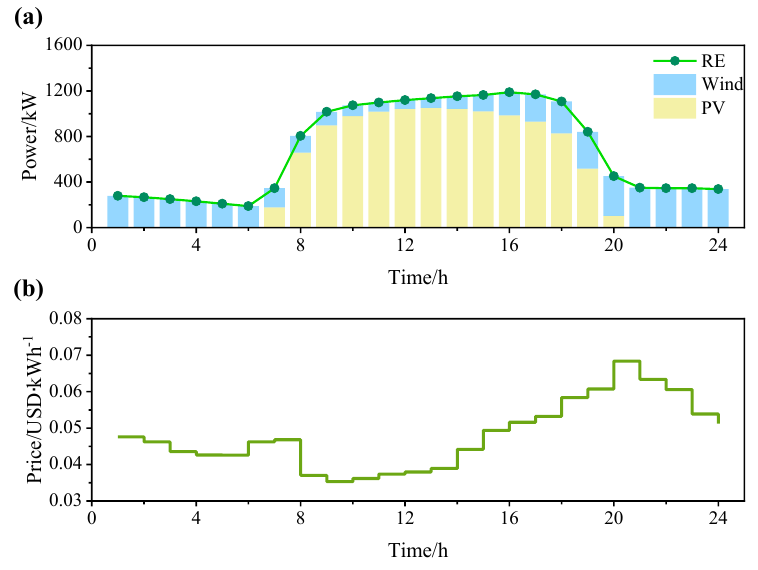}
\caption{Case-study input: (a) forecasting output of wind and solar generation; (b) time-of-use electricity price.}
\label{fig:case_inputs}
\end{figure}

Considering a green-power direct-supply AIDC with 2048 GPUs, we assume that this AIDC is only used for LLM pre-training tasks. Renewable-generation profiles and electricity prices are selected from CAISO. The profiles are matched to the case capacities in Table~\ref{tab:power_params}. Fig.~\ref{fig:case_inputs} shows the renewable and TOU-price inputs, where fixed continuous training is not naturally aligned with renewable availability or electricity price.

\begin{table}[!t]
\caption{Representative LLM Pre-training Parameters}
\label{tab:task_params}
\centering
\footnotesize
\setlength{\tabcolsep}{2pt}
\renewcommand{\arraystretch}{0.94}
\begin{tabular}{p{0.38\columnwidth}>{\centering\arraybackslash}p{0.27\columnwidth}>{\centering\arraybackslash}p{0.27\columnwidth}}
\toprule
Parameter & LLM-A & LLM-B \\
\midrule
Representative model & LLaMA-13B & LLaMA-33B \\
Model size/training tokens & 13.0B/1.0T & 32.5B/1.4T \\
Layers/$h$/$h_f$/heads & 40/5120/13824/40 & 60/6656/17920/52 \\
Sequence length/vocabulary & 2048/32000 & 2048/32000 \\
Reported GPU-hours & 135168 & 530432 \\
Baseline $(p,t,d)$/GPUs & $(4,4,32)$/512 & $(6,4,32)$/768 \\
Daily ratio & 4.900\% & 2.021\% \\
Segments/deadline & 12/24 h & 12/24 h \\
\bottomrule
\end{tabular}
\end{table}

We choose two representative LLaMA pre-training tasks, as showen in Table~\ref{tab:task_params} . Model architecture, token targets, nominal batch, and GPU-hours follow public data \cite{Touvron2023LLaMA}. The implemented batch, baseline modes, checkpoint segments, and daily tranches are case settings; FFN dimensions and daily compute are derived. Table~\ref{tab:power_params} lists the case designed capacities and costs. The AIDC load follows from the base load, GPU power, pool size, and PUE \cite{Barroso2019Datacenter}.

\begin{table}[!t]
\caption{Key AIDC and Power-System Parameters}
\label{tab:power_params}
\centering
\footnotesize
\setlength{\tabcolsep}{2pt}
\renewcommand{\arraystretch}{0.98}
\begin{tabular}{p{0.25\columnwidth}>{\centering\arraybackslash}p{0.20\columnwidth}p{0.24\columnwidth}>{\centering\arraybackslash}p{0.23\columnwidth}}
\toprule
Parameter & Value & Parameter & Value \\
\midrule
GPU type & A100 & Peak FLOPS/GPU & 312 TFLOPS \\
Physical (GB) & 80 & PUE & 1.25 \\
Base IT load (kW) & 375 & Power/GPU (kW) & 0.40 \\
PV/wind (kW) & 1050/350 & BESS kW/kWh & 300/800 \\
BESS efficiency & 0.95 & Curt./BESS cost & 20.83USD/MWh \\
\bottomrule
\end{tabular}
\end{table}

\subsubsection{Validation Scenarios}
Six scenarios are designed to isolate the effects of checkpoint information, 3D-mode variability, and BESS operation, as shown in Table~\ref{tab:scenarios}.

\begin{table}[!t]
\caption{Scenario Definitions}
\label{tab:scenarios}
\centering
\scriptsize
\resizebox{1\columnwidth}{!}{%
\begin{tabular}{lccc}
\toprule
Scenario & BESS & Checkpoint & 3D modes \\
\midrule
S1: Fixed-NoCkpt-NoBat & No & No & Fixed \\
S2: CkptRecord-Fixed3D-NoBat & No & Yes & Fixed \\
S3: 3D-NoCkpt-NoBat & No & No & Variable \\
S4: Ckpt-3D-NoBat & No & Yes & Variable \\
S5: Fixed-Bat & Yes & No & Fixed \\
S6: Ckpt-3D-Bat & Yes & Yes & Variable \\
\bottomrule
\end{tabular}%
}
\end{table}

S1 is the fixed baseline. S2 only records checkpoint states along the S1 trajectory and does not allow waiting, so S1 and S2 are intentionally identical. S3 is an idealized 3D-only benchmark without checkpoint continuity. S4 is the executable checkpoint-aware 3D case, S5 isolates BESS under fixed training, and S6 is the full coordination case.

\subsection{Overall Scenario Comparison}

\begin{table*}[!t]
\caption{Overall Scenario Performance}
\label{tab:baseline_results}
\centering
\scriptsize
\resizebox{\textwidth}{!}{%
\begin{tabular}{lcccccccccc}
\toprule
\multirow{2}*{Scenario} & \multirow{2}*{Main mechanism} & Cost & Cost red. & Grid & Curt. & BESS thru. & RE util. & Green share & Peak power & Peak GPU \\
 & & (USD) & (\%) & (MWh) & (MWh) & (MWh) & (\%) & (\%) & (kW) & \\
\midrule
S1 & Fixed baseline & 473.03 & 0.00 & 8.965 & 1.077 & 0.000 & 93.46 & 63.20 & 1015.0 & 1280 \\
S2 & Checkpoint record & 473.03 & 0.00 & 8.965 & 1.077 & 0.000 & 93.46 & 63.20 & 1015.0 & 1280 \\
S3 & Idealized 3D only & 368.99 & 21.99 & 8.395 & 0.000 & 0.000 & 100.00 & 66.24 & 1399.0 & 2048 \\
S4 & Checkpoint+3D & 378.33 & 20.02 & 8.434 & 0.014 & 0.000 & 99.92 & 66.12 & 1335.2 & 2048 \\
S5 & Fixed+BESS & 447.23 & 5.45 & 8.281 & 0.319 & 1.442 & 98.06 & 66.00 & 1015.0 & 1280 \\
S6 & Checkpoint+3D+BESS & 377.79 & 20.13 & 8.422 & 0.000 & 0.026 & 100.00 & 66.17 & 1335.2 & 2048 \\
\bottomrule
\end{tabular}
}
\end{table*}

Table~\ref{tab:baseline_results} compares the six scenarios. S1 is the fixed continuous-training baseline, whereas S2 only records checkpoint states along the same fixed trajectory. Their identical results show that checkpoint information alone cannot reshape the load unless checkpoint waiting and subsequent recovery actions are enabled. S3 achieves the lowest cost among the no-storage cases, reducing cost by 21.99\% and eliminating curtailment. However, it allows idealized 3D-mode variation without enforcing checkpoint-state continuity and should therefore be interpreted as an optimistic benchmark rather than a physically executable policy.

In contrast, S4 enforces checkpoint-state continuity and still reduces cost by 20.02\% while lowering renewable curtailment from 1.077 to 0.014~MWh relative to S1. Its cost is only 9.34~USD above that of S3, so it retains approximately 91\% of the idealized cost saving under explicit execution constraints. The storage comparison further clarifies the source of flexibility. S5 reduces cost by only 5.45\% and requires 1.442~MWh of BESS throughput because the training trajectory remains fixed. S6 combines checkpoint-aware 3D dispatch with BESS, reducing cost by 20.13\%, eliminating curtailment, and requiring only 0.026~MWh of BESS throughput. The additional cost reduction from S4 to S6 is only 0.14\%, indicating that executable workload reshaping provides the dominant benefit, while BESS corrects the small residual source--load mismatch.

\subsection{Source--Load--Storage--Grid Dispatch Comparison}

\begin{figure}[!t]
\centering
\includegraphics[width=\columnwidth]{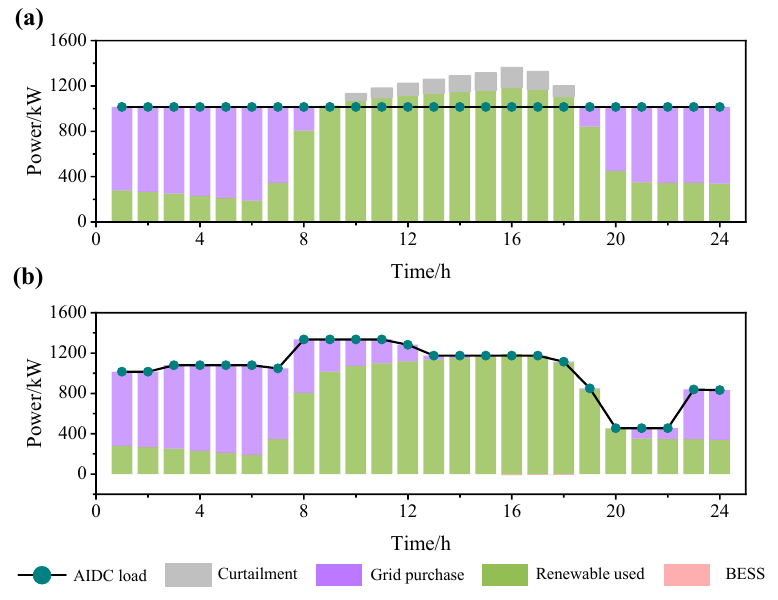}
\caption{Source--load--storage--grid balance of representative scenarios: (a) S1 fixed continuous training; (b) S6 checkpoint-aware 3D dispatch with BESS.}
\label{fig:dispatch_balance}
\end{figure}

Fig.~\ref{fig:dispatch_balance} compares the 24-h power balance of S1 and S6. In S1, the two LLM profiles follow fixed continuous-training trajectories, and the AIDC load therefore remains at 1015~kW throughout the day. This rigid demand cannot increase to absorb renewable surpluses or decrease when renewable output falls and the electricity price rises. Consequently, S1 purchases 8.965~MWh from the grid and curtails 1.077~MWh of renewable energy, with a renewable utilization rate of 93.46\% and a green-power share of 63.20\%.

In S6, checkpoint waiting and feasible 3D-mode selection reshape the AIDC load between 455.0 and 1335.2~kW. The scheduler raises training demand during renewable-rich periods and lowers it in the late evening, while BESS performs only a brief residual charge--discharge adjustment. This coordination eliminates renewable curtailment, reduces grid purchase from 8.965 to 8.422~MWh, and increases the green-power share from 63.20\% to 66.17\%. Thus, the proposed method does not simply reduce electricity consumption; it redistributes executable training work in time to align demand with renewable availability, with storage serving as a complementary balancing resource.

\subsection{Mechanism Analysis of Checkpoint-Aware 3D Training Dispatch}

\begin{figure}[!t]
\centering
\includegraphics[width=\columnwidth]{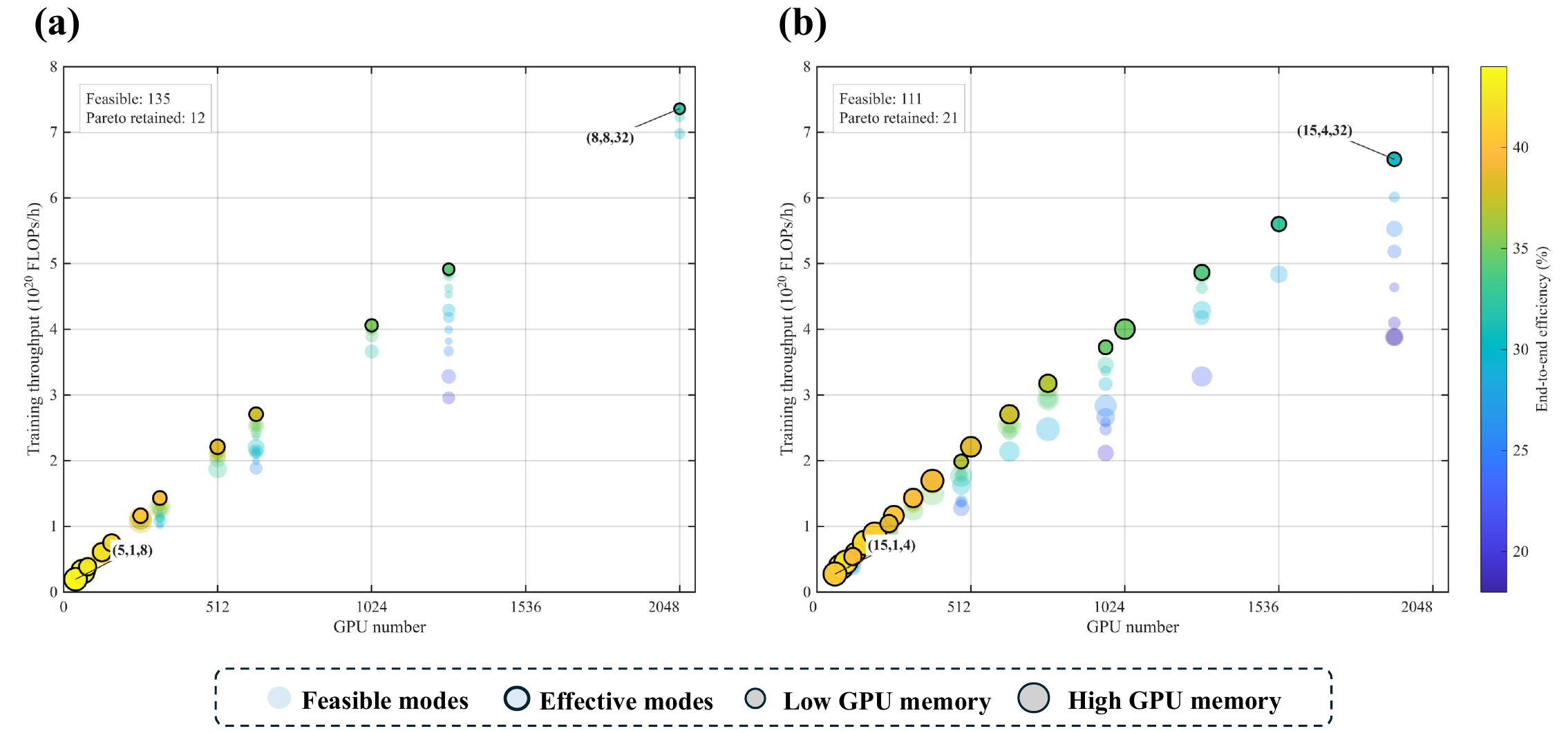}
\caption{Feasible 3D-parallel configuration space and retained effective modes: (a) LLM-A-13B; (b) LLM-B-33B.}
\label{fig:config_space}
\end{figure}

\begin{figure}[!t]
\centering
\includegraphics[width=\columnwidth]{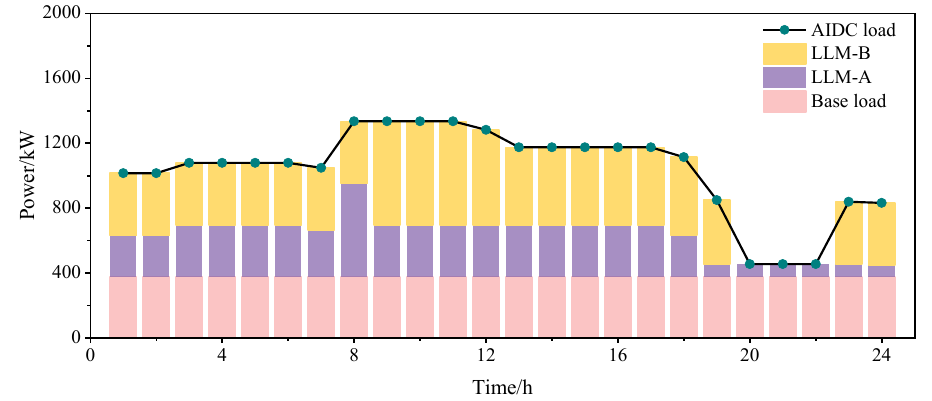}
\caption{AIDC load composition under the full S6 dispatch model.}
\label{fig:load_composition}
\end{figure}

\begin{figure}[!t]
\centering
\includegraphics[width=\columnwidth]{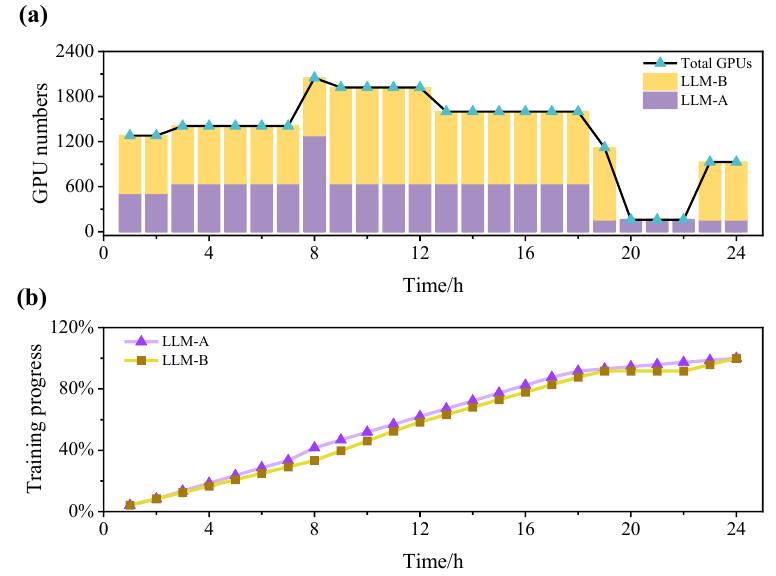}
\caption{GPU allocation and daily training progress under the full S6 dispatch model: (a) GPU usage; (b) training progress.}
\label{fig:gpu_progress}
\vspace{-3pt}
\end{figure}

Fig.~\ref{fig:config_space} shows the discrete operating modes available to the scheduler. After memory, divisibility, and GPU-capacity screening, 135 feasible configurations are generated for LLM-A and 111 for LLM-B. Pareto pruning retains 12 and 21 effective modes, respectively, whereas the optimal S6 schedule activates only four modes for LLM-A and three for LLM-B. These retained modes provide distinct GPU--throughput--power combinations that allow delayed training work to be recovered without introducing arbitrary continuous load scaling or carrying the full configuration space into the dispatch problem.

Figs.~\ref{fig:load_composition} and \ref{fig:gpu_progress} verify that the resulting electrical trajectory is execution-feasible. GPU allocation reaches the 2048-GPU cluster limit during recovery, and the corresponding AIDC power varies from 455.0 to 1335.2~kW. LLM-A does not pause and continues training in a 160-GPU low-power mode during the late-evening interval, whereas LLM-B waits at a checkpoint for three hours and resumes at hour 22. Its cumulative progress remains unchanged during waiting and subsequently recovers through feasible 3D modes. Both profiles complete 100\% of their daily compute by hour 24 with zero boundary idle. The load steps in Fig.~\ref{fig:load_composition} therefore correspond to explicit checkpoint and 3D-mode decisions, confirming that the reported power-system benefits are obtained without violating training progress or execution constraints.

\subsection{Sensitivity Analysis}

\begin{figure*}[!t]
\centering
\includegraphics[width=\textwidth]{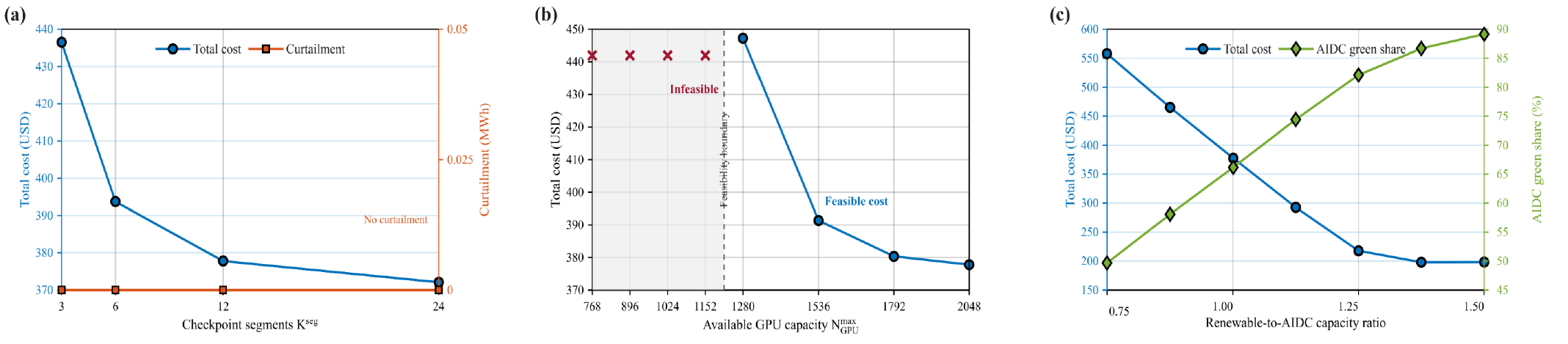}
\caption{Sensitivity analysis of key dispatch parameters: (a) checkpoint granularity; (b) GPU capacity; (c) renewable integration.}
\label{fig:sensitivity}
\vspace{-3pt}
\end{figure*}

Fig.~\ref{fig:sensitivity}(a) shows diminishing checkpoint value. The final cost improvement is only 1.51\%; 12 checkpoints segments capture most of the available flexibility. Fig.~\ref{fig:sensitivity}(b) identifies a GPU feasibility threshold. Capacities of 768--1152 GPUs cannot complete both tranches, while 1280 GPUs restore feasibility at 447.23~USD. Most further savings are obtained by 1536 GPUs, after which cost approaches 377.79~USD at 2048. GPUs are therefore most valuable when they first enable post-waiting recovery. In Fig.~\ref{fig:sensitivity}(c), increasing the renewable-to-AIDC ratio from 0.75 to 1.375 lowers cost from 557.84 to 197.94~USD and raises green share from 49.66\% to 86.69\%. At 1.5, green share reaches 89.14\%, but cost rises slightly and curtailment reaches 2.127~MWh. This reversal marks the point at which additional renewable energy exceeds the absorption capability of workload and storage flexibility.

\section{Conclusion}
This paper proposed an execution-feasible dispatch framework for representative LLM pre-training load profiles in a grid-connected green-power direct-supply AIDC. The proposed framework maps LLM descriptors to checkpoint-domain workloads and feasible 3D-parallel GPU modes, and embeds the retained execution modes into a checkpoint-state graph coupled with renewable curtailment, grid purchase, and BESS operation.

Numerical studies on a 2048-GPU AIDC with two representative LLaMA-derived profiles show that checkpoint-aware 3D dispatch reduces the operating cost by 20.02\% and decreases renewable curtailment from 1.077~MWh to 0.014~MWh compared with fixed continuous training. When BESS is further coordinated, the full dispatch case reduces the operating cost by 20.13\%, eliminates renewable curtailment, and requires only 0.026~MWh of BESS throughput. The results confirm that checkpoint information alone does not create load flexibility; useful flexibility appears only when checkpoint waiting is combined with feasible 3D-mode compensation. In the studied case, workload reshaping provides the dominant flexibility, while BESS performs residual balancing.

Note that the numerical profiles and generated 3D modes are constructed from public trace and data, rather than proprietary production traces. Future work will extend the framework to multi-day stochastic dispatch and other AI workloads such as fine-tuning and inference, include idle-GPU opportunity cost under shared GPU pools, and investigate coordinated dispatch among multiple AIDCs.

\balance

\end{document}